\documentclass[a4paper]{article}
\usepackage{epsfig}
\title{CP Violation in Seesaw Model}
\renewcommand\epsilon{\varepsilon}
\begin{document}
\author{T. Endoh\thanks{E-mail:endoh@theo.phys.hiroshima-u.ac.jp},
 T. Morozumi\thanks{E-mail:morozumi@theo.phys.sci.hiroshima-u.ac.jp},
 T. Onogi\thanks{E-mail:onogi@theo.phys.sci.hiroshima-u.ac.jp} and
 A. Purwanto\thanks{E-mail:purwanto@theo.phys.sci.hiroshima-u.ac.jp.On
 leave from Jurusan Fisika FMIPA-ITS, Surabaya Indonesia. }}
\date{\small \it Department of Physics, Hiroshima University \\
        1-3-1 Kagamiyama, Higashi Hiroshima - 739-8526, Japan}
\maketitle
\thispagestyle{empty}
\begin{abstract}
We study the structure of CP violating phases in the seesaw model. 
We find that the $3\times 6$ MNS matrix contains six independent
phases, three of which are identified as a Dirac phase and two 
Majorana phases in the light neutrino sector 
while the remaining three arise from the mixing of 
the light neutrinos and heavy neutrinos.
We show how to determine these phases from physical observables.
\end{abstract}
\clearpage
\clearpage
\section{Introduction}
The recent experiment on neutrino oscillations~\cite{Fu} strongly suggests the
nonzero masses of neutrinos. Among all the models of neutrino masses, 
one of the most attractive ways to generate the smallness of neutrino 
masses is the seesaw mechanism~\cite{YaGe}, which introduces
right-handed massive neutrinos. Since the right-handed neutrinos are
so heavy, we usually integrate them out and consider the low energy 
effective theory with only the left-handed neutrinos. However, 
it is possible that the low energy effective theory might miss
some of the essential features of CP violation in the seesaw model. 
Therefore it is important to study the general structure of 
the phases in the mixing matrix in the seesaw model 
which give the CP violation and the lepton number asymmetry. 

When we integrate out the heavy mode and consider the low energy
effective theory, which we call as ``decoupling
case'' throughout this paper, all the CP violating phases on the
lepton sector are contained in the unitary 3 ${\times}$ 3 MNS mixing
matrix~\cite{MaSa}. On the other hand, when we consider the full 
theory of seesaw model with the heavy right-handed neutrinos, 
which we call as ``non-decoupling case'', 
the MNS matrix is 3 ${\times}$ 6 matrix. The
goal of this paper is to understand degrees of freedom of independent CP
violating phases contained in the 3 ${\times}$ 6 MNS matrix in the
non-decoupling case, understand where the difference from the
conventional decoupling case arises, and show how the phase
including can be determined from experimental observation.

This paper is organized as follows.
In section 2, we introduce the extended MNS matrix and 
count the physical degrees of freedom of the matrix 
which remains after imposing unitarity conditions and seesaw 
conditions and using the rephasing symmetry from which 
we find that there exist six independent phases. 
As an explicit example, we confirm
this result in a special case where the MNS matrix can be
explicitly parametrized.
In section 3 we study how to determine the
phase of the MNS matrix and show that one phase is sensitive in
neutrino oscillations, two phases are sensitive in neutrinoless double
beta decay and the other three phases are sensitive in lepton number
asymmetry. We also present geometrical interpretation of the
effects of the phases hidden in Yukawa coupling. Finally, we discuss and
conclude the results of our work in section 4. 

\section{MNS matrix}
The Lagrangian of the seesaw model is given as follows 
\begin{equation}
{\cal L}=-y_{\nu}^{ij}\overline{{\psi}_{L}}^{i}\tilde{\phi}N_{R}^{j}-\frac{1}{2}\overline{(N_{R}^{0})^{c}}m_{N}N_{R}^{0}-y_{e}^{ij}\overline{{\psi}_{L}}^{i}{\phi}e_{R}^{j}+h.c., 
\label{1}
\end{equation} 
where $\tilde{\phi}=i{\tau}^{2}{\phi}^{*}$. Without loss of generality we can start with real diagonal matrix
for $m_N$. This Lagrangian gives mass term as 
\begin{equation}
{\cal L}_{M}=-\frac{1}{2}(\overline{\nu}_{L}^{0}\
\overline{(N_{R}^{0})^{c}})M_{\nu}\left(
\begin{array}{c}
({\nu}_{L}^{0})^{c} \\
N_{R}^{0} 
\end{array}
\right)+h.c.,
\label{2}
\end{equation} 
where the mass matrix $M_{\nu}$ is a symmetric matrix given as follows,
\begin{equation}
M_{\nu}=\left(\begin{array}{cc}
0 & Y_{\nu}\frac{v}{\sqrt{2}} \\
Y^{T}_{\nu}\frac{v}{\sqrt{2}} & m_{N}
\end{array}
\right). 
\label{3}
\end{equation} 
where we define a matrix as $[Y_{\nu}]_{ij}=y_{\nu}^{ij}$. Hence, this mass matrix can be diagonalized by unitary matrix V, i.e.
$m_d=V^{T}M_{\nu}V$, where $V$ is mixing matrix,
\begin{equation}
\left(\begin{array}{c}
{\nu}_{L}^{0} \\
(N_{R}^{0})^{c}\end{array}
\right)=V^{*}\left(\begin{array}{c}
{\nu}_{L}^{\alpha}
\end{array}
\right). 
\label{4}
\end{equation} 

The charged current ${J^{\dagger}}^{\mu}=
\overline{l}_{L}^{i}V_{MNS}^{i{\alpha}}{\gamma}^{\mu}{\nu}_{L}^{\alpha}$, 
where indices i = 1,2,3 and ${\alpha}$=1,2,..,6. Since we can always start with mass eigenstate for charged lepton, then the MNS matrix is nothing but the mixing matrix $V$ itself,

\begin{equation}
V_{MNS}^{i{\alpha}}=V^{i{\alpha}}. 
\label{5}
\end{equation} 
It's clear from (\ref{5}) that $V_{MNS}$ is $3{\times}6$ matrix.\\
We now count the physical degrees of freedom contained in this complex 
matrix $V_{MNS}$ . For N generations, $V_{MNS}$ is $N{\times}2N$ complex matrix. Hence, it has $2N^
2$ real and imaginary parts respectively. This matrix satisfies
two conditions. First is the unitarity condition which is given in the form below,  
\begin{equation}
V_{MNS}V_{MNS}^{\dagger}={\bf 1_{3}}. 
\label{6}
\end{equation} 
${\bf 1_{3}}$ is 3${\times}$3 unit matrix. The unitarity condition gives
$\frac{N^{2}+N}{2}$ constraints for real parts and 
$\frac{N^{2}-N}{2}$ for imaginary parts, i.e., phases.
The second is the very special condition coming from the seesaw type mass
matrix. We call it {\it seesaw condition} from zeros of the matrix
(\ref{3}) and the diagonalization leads to
\begin{equation}
\biggl(V_{MNS}( m_d )V_{MNS}^{T}\biggr)^{ij}=0. 
\label{7}
\end{equation} 
The seesaw condition gives $\frac{N^{2}+N}{2}$ constraints for real parts
and $\frac{N^{2}+N}{2}$ for phases. By taking into
account the unitarity and seesaw conditions, there remain  $N^2-N$ real
parts and $N^2$ phases. Finally, after absorbing N
unphysical phases into the charged lepton fields we have $N^2-N$
independent physical phase. The total independent parameters are
summarized in table below.
\begin{center}
Table I
\end{center}
$$
\begin{array}{|c|c|c|c|c|c|}\hline
\multicolumn{2}{|c|}{\bf {V_{MNS}}} &\multicolumn{3}{|c|}{\mbox{\bf Constraints}} &  \mbox{\bf Independent parameter} \\ \hline
\multicolumn{2}{|c|}{[N{\times}2N]} & \mbox{unitarity} & \mbox{Seesaw } & \mbox{Rephasing} & \\ \hline 
\mbox{Real} & 2N^{2} & N+\frac{N^{2}-N}{2} & \frac{N^{2}+N}{2} & &
N^{2}-N \\ \hline 
\mbox{Im} & 2N^{2} & \frac{N^{2}-N}{2} & \frac{N^{2}+N}{2} & N & N^{2}-N
\\ \hline 
\mbox{Total} & 4N^{2} & N^{2} & N^{2}+N & N & 2N(N-1) \\ \hline
\end{array}
$$
For $N=3$, the number of independent phases is $6$. The result is different
from which could be obtained the one by integrating out the heavy
neutrino, i.e. one phase for Dirac neutrino and three phase for
Majorana neutrino. We will show how three extra phases come from.

We confirm this result explicitly by taking a specific example. 
Consider the case where the majorana masses in the heavy right-handed 
mass has a strong hierarchy in the diagonal basis, namely 
$m_{N1} \gg m_{N2} \gg m_{N3} ( \gg v )$,
in which case, the approximate diagonalization of the seesaw matrix is possible~\cite{MoTa,HaMo}. The idea is that starting from the decoupling limit,
the mass matrix in Eq.(\ref{3}) can be diagonalized using a 
systematic expansion in $v/m_N$.
As a result,
 the $V_{MNS}$ is given by 
\begin{equation}
V_{MNS}= \Biggl(U\ \ {Y_{\nu}\frac{v}{\sqrt{2}m_{N}}}_{\mbox{}}\Biggr), 
\label{8}
\end{equation} 
where $U$ is a unitary matrix
which transforms $Y_{\nu}$ into triangular matrix $Y_{\triangle}$,  
\begin{equation}
Y_{\nu} = {\bf U}Y_{\triangle}= {\bf U}\left(\begin{array}{ccc}
y_{1} & 0 & 0 \\
y_{21} & y_{2} & 0 \\
y_{31} & y_{32} & y_{3} 
\end{array}\right). 
\label{9}
\end{equation} 
(See \cite{MoTa} and Appendix A for the proof.)
The diagonal elements of $Y_{\triangle}$ are real. The general $3{\times}3$ Yukawa coupling has nine phases. The
decomposition shows that six of them are included in $U$ while the other
three are included in the off-diagonal elements of
$Y_{\triangle}$. Using the decomposition, we can rewrite $V_{MNS}$ as
\begin{equation}
V_{MNS}=U\left({\bf 1}_3,\ Y_{\triangle}\frac{v}{\sqrt{2}m_{N}}\right).
\label{10}
\end{equation}
Three of six phases in $U$ can be absorbed into the definition
of charged leptons. Therefore we conclude that $V_{MNS}$ has six phases
 in total, three of them are in $U$ and the other three are in
$Y_{\triangle}$. 
This is consistent with
the results shown in Table I. The $N{\times}N$ triangular
matrix $Y_{\triangle}$ with real diagonal elements
includes $\frac{N^{2}-N}{2}$ phases.
This is exactly the difference of the number of independent phases
between non-decoupling case and decoupling case.
We call these phases as heavy phase.
Then, in general $N(N-1)$ independent phase can be separated as follows
\begin{equation}
N(N-1)=\underbrace{\underbrace{\frac{(N-2)(N-1)}{2}}_{\mbox{\tiny  Dirac phase}}+\underbrace{(N-1)}_{\mbox{\tiny  Majorana phase}}}_{\mbox{\tiny Decoupling case}}+\underbrace{\frac{(N^{2}-N)}{2}}_{\mbox{\tiny heavy}}. 
\label{11}
\end{equation} 
What we call heavy phase in (\ref{11}) correspond to the phases
included in $Y_{\triangle}$, i.e., the extra phases come from 
off-diagonal elements of
$(Y_{\triangle})_{ij}\ \mbox{for}\ i > j$. The numbers of CP
violating phases for $N=2$ and $N=3$
are given in the following

\vspace{1cm}
\begin{center}
Table II
\end{center}
$$
\begin{array}{|c|c|c|c|}\hline
 & \mbox{Dirac} & \mbox{Majorana} & \mbox{heavy}\\ \hline
N=2 & 0 & 1 &  1 \\ \hline
N=3 & 1 & 2 &  3 \\ \hline
\end{array}
$$

\section{CP Violation}
Now that we know the number of independent phase of the $V_{MNS}$
matrix, the question is how we can determine these independent phases. 
To answer this question we discuss the neutrino oscillations,
neutrinoless double beta decay and lepton number asymmetry as phenomena which
 are sensitive to the phases of the extended MNS matrix discussed in
the previous sections. This extended MNS matrix which relates ${\nu}_{L}^0$ and ${\nu}_{L}$ as Eq.(\ref{4}), i.e.
${\nu}_{L}^{0i} = V_{MNS}^{*i{\alpha}}{\nu}_{L}^{\alpha}$ gives the
extended amplitude for both neutrino oscillations and neutrinoless
double beta decay.

First, we consider the neutrino oscillations. The transition amplitude ${\nu}_{i}\ {\rightarrow}\
{\nu}_{j}$ is extended from the decoupling case as follows
\begin{eqnarray}
A_{{\nu}_{i}{\rightarrow}{\nu}_{j}}(t)= {\langle}{\nu}_{j}|{\nu}_{i}(t){\rangle}
& = &
\sum_{k}U_{jk}^{*}U_{ik}e^{-iE_{k}t}+\sum_{k}Y_{jk}^{*}Y_{ik}\frac{v^{2}}{2{m_{Nk}^{2}}}e^{-iE'_{k}t}\nonumber \\ 
&{\simeq}& \sum_{k}U_{jk}^{*}U_{ik}e^{-iE_{k}t}.
\label{12}
\end{eqnarray} 
In the last expression, we dropped the contributions
from the massive neutrino since they are tiny so that 
we obtain the same result as in the decoupling case.
Hence, we obtain the asymmetry~\cite{AraSa}  
\begin{equation}
{\eta}=P_{{k}{\rightarrow}{l}}-P_{\overline{k}{\rightarrow}\overline{l}}=2\sum_{ij}\mbox{Im}\Bigl\{U_{CKM}^{*li}({\theta})U_{CKM}^{lj}({\theta})U_{CKM}^{ki}({\theta})U_{CKM}^{*kj}({\theta})\Bigr\}\sin\Biggl({\delta}m_{ij}^{2}\frac{L}{2E}\Biggr), 
\label{13}
\end{equation} 
since the unitary matrix U can always be transformed $U {\rightarrow} U_{CKM}({\theta})\mbox{diag}(1, e^{i{\zeta}_{1}}, e^{i{\zeta}_{2}})$,
where ${\theta}$ is Dirac phase and ${\zeta}_1, {\zeta}_2$ are Majorana
phases (see Appendix A).
Eq.(\ref{13}) represents the CP violation in neutrino oscillations and we
can see that neutrino oscillation is only sensitive on Dirac phase living in CKM matrix but not on Majorana phase ${\zeta}_{1}$ and ${\zeta}_{2}$. 

The similar is also the case in neutrinoless double beta decay. The
amplitude is modified from the decoupling case due to the 
contribution from the heavy neutrino, but after
taking into account the scale of momentum-energy and the neutrino
masses, the expansion in components may be reduced to 
the same form as in the decoupling case~\cite{DoiTak} 
\begin{eqnarray}
\sum_{\alpha}\frac{m_{\alpha}}{q^{2}+m_{\alpha}^{2}}V_{e{\alpha}}^{2}& = &
\frac{1}{q^{2}}\sum_{i}m_{i}U_{ei}^{2}+\frac{v^{2}}{2}\sum_{i}\frac{1}{m_{N_{i}}^{3}}Y_{ei}^{2}\nonumber \\ 
& {\simeq} & \frac{1}{q^{2}}\sum_{i}m_{i}U_{ei}^{2}({\theta},{\zeta}_{1},{\zeta}_{2}). 
\label{14} 
\end{eqnarray} 
It is seen that in principle all phases of low energy sector can be
determined by using double beta decay experiments, but not for the 
remaining three phases hidden in Yukawa coupling.

Finally, we consider the lepton number asymmetry in the seesaw model, 
which was proposed as one of the most promising scenarios 
of baryogenesis~\cite{FuYa}. The lepton number asymmetry has been 
studied in the flavor basis. Here we rederive the formula for 
lepton number asymmetry in the mass basis, which is another important
result of the present paper. In this basis, the lepton number
asymmetry is expressed in terms of only the physical quantities
such as mass and $V_{MNS}$ matrix. Therefore in our result 
the relation of lepton number asymmetry and the CP violating phases 
is obvious. The relevant part of the Lagrangian which may be derived
from Eq.(\ref{1}) is\footnote{The process corresponds to $N^{\alpha}\
{\rightarrow}\ l^{\pm}W^{\mp}$. It can be shown that the decay into the
longitudinal $W$ mesons can be approximately described by the decay
into the unphysical Goldstone boson.}
\begin{equation}
{\cal L}_{\chi}=\frac{\sqrt{2}}{v}{\chi}^{+}\overline{({\nu}_{L}^{c})}^{\alpha}m^{\alpha}(V_{MNS}^{\dagger})^{{\alpha}k}l_{L}^{k}+h.c. 
\label{15}
\end{equation} 
From Eq.(\ref{15}) we get the decay amplitudes ${\cal M}$,
\begin{equation}
\begin{array}{l}
{\displaystyle {\cal M}^{tree}= \Biggl(\frac{\sqrt{2}m_{\alpha}}{v}\Biggr)(V^{i{\alpha}}_{MNS})^{*}\Bigl(\overline{U_{e}}LU_{N}\Bigr)}.  \\
{\displaystyle {\cal M}^{v}= -i\Biggl(\frac{\sqrt{2}m_{\alpha}}{v}\Biggr)\Biggl(\frac{\sqrt{2}m_{\beta}}{v}\Biggr)^{2}V^{j{\alpha}}_{MNS}(V_{MNS}^{i{\beta}})^{*}(V^{j{\beta}}_{MNS})^{*}\frac{I(x)}{16{\pi}}\Bigl(\overline{U_{e}}LU_{N}\Bigr).} \\
{\displaystyle {\cal M}^{s}= -i\Biggl(\frac{\sqrt{2}m_{\alpha}}{v}\Biggr)\Biggl(\frac{\sqrt{2}m_{\beta}}{v}\Biggr)^{2}\frac{m_{\alpha}m_{\beta}}{{m_{\alpha}}^{2}-{m_{\beta}}^{2}}V_{MNS}^{j{\alpha}}(V_{MNS}^{j{\beta}})^{*}(V_{MNS}^{i{\beta}})^{*}\frac{1}{32{\pi}}\Bigl(\overline{U_{e}}LU_{N}\Bigr).} \\ 
\end{array}
\label{16}
\end{equation}
where the upper indices show the tree, vertex and self-energy correction
respectively, whereas $U_{e},\ U_{N}$ are spinors of electron and right handed
Majorana neutrino, and $I(x)$ is
\begin{equation}
I(x)=\sqrt{x}\Biggl[1+(1+x)log\Biggl(\frac{x}{1+x}\Biggr)\Biggr].\label{36} 
\end{equation} 
with  $x = ({m_{\beta}^{2}}/{m_{\alpha}^{2}})$. For
one-loop amplitudes ${\cal M}^{v}$ and ${\cal M}^{s}$, we have evaluated the absorptive
part only which contributes to the lepton number asymmetry
through the interference with tree level amplitude ${\cal M}^{tree}$.  

The interference between tree level and one-loop correction of the
heavy Majorana neutrino decay gives the asymmetry of the lepton number
\begin{equation}
a(N_{\alpha}{\rightarrow}{l_i^{\mp}{\chi}^{\pm}}) = a_{{\alpha}i}= 
\displaystyle{\frac
{{\Gamma}(N_{\alpha}{\rightarrow}l^{-}_i{\chi}^{+})-{\Gamma}(N_{\alpha}{\rightarrow}l^{+}_i{\chi}^{-})}{\sum_{i}{\Gamma}(N_{\alpha}{\rightarrow}l^{-}_i{\chi}^{+})+{\Gamma}(N_{\alpha}{\rightarrow}l^{+}_i{\chi}^{-})}} 
\label{18}
\end{equation}

The contribution from the interference of tree and vertex diagram is
given as

\begin{figure}[htbp]
\begin{tabular}{ccc}
\begin{minipage}{0.3\hsize}
\begin{center}
\epsfig{file=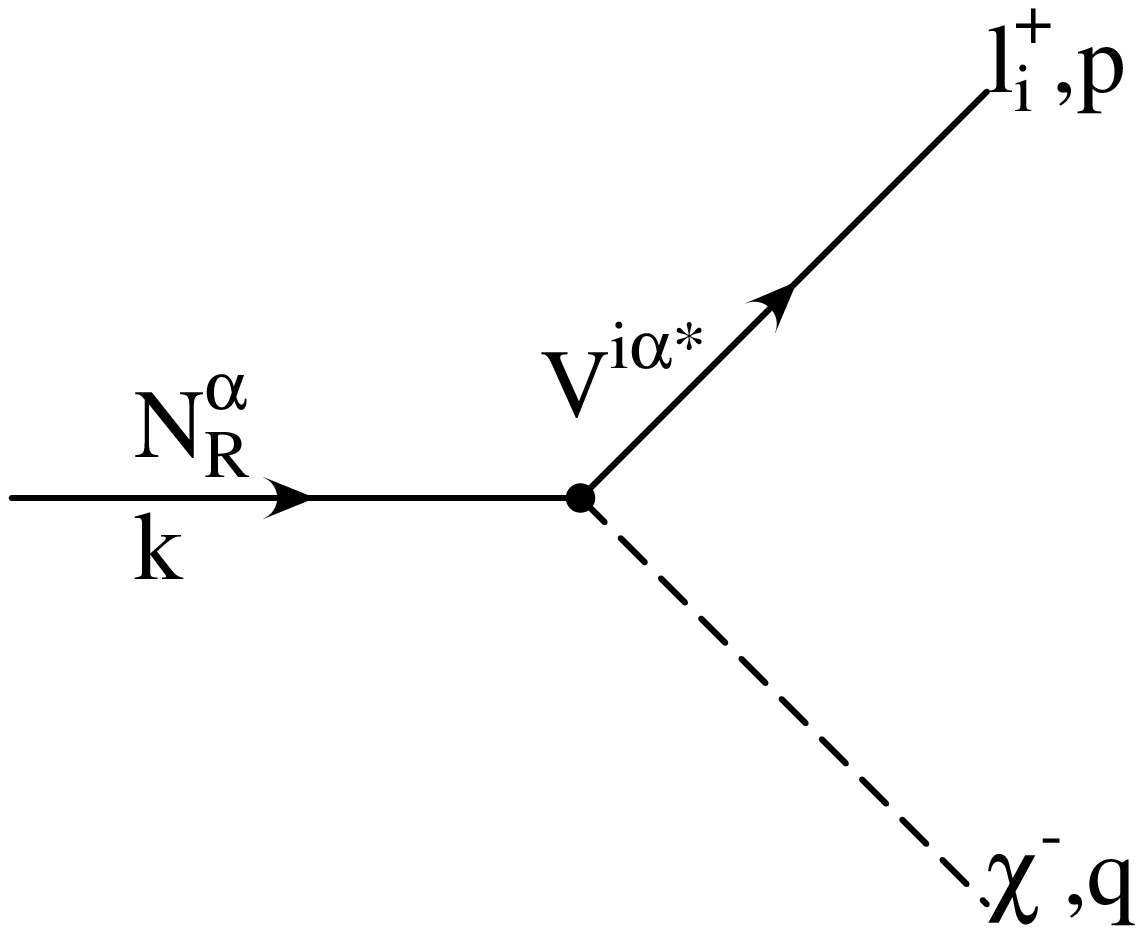,width=\hsize}
\label{fig:Tree}
\end{center}
\end{minipage}
\begin{minipage}{0.2\hsize}
\begin{center}
\epsfig{file=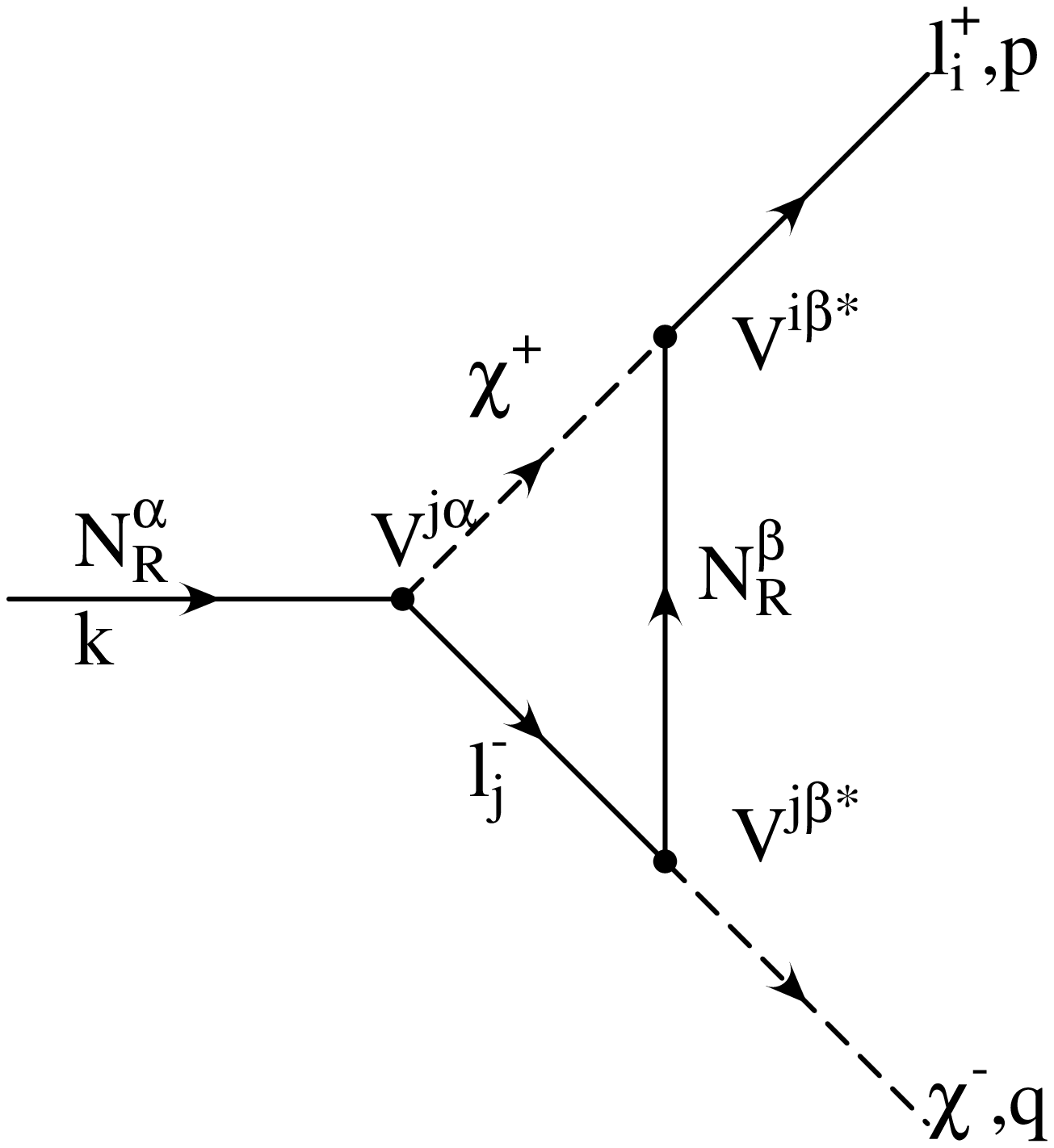,width=\hsize}
\label{fig:1loop}
\end{center}
\end{minipage}
\begin{minipage}{0.3\hsize}
\begin{center}
\epsfig{file=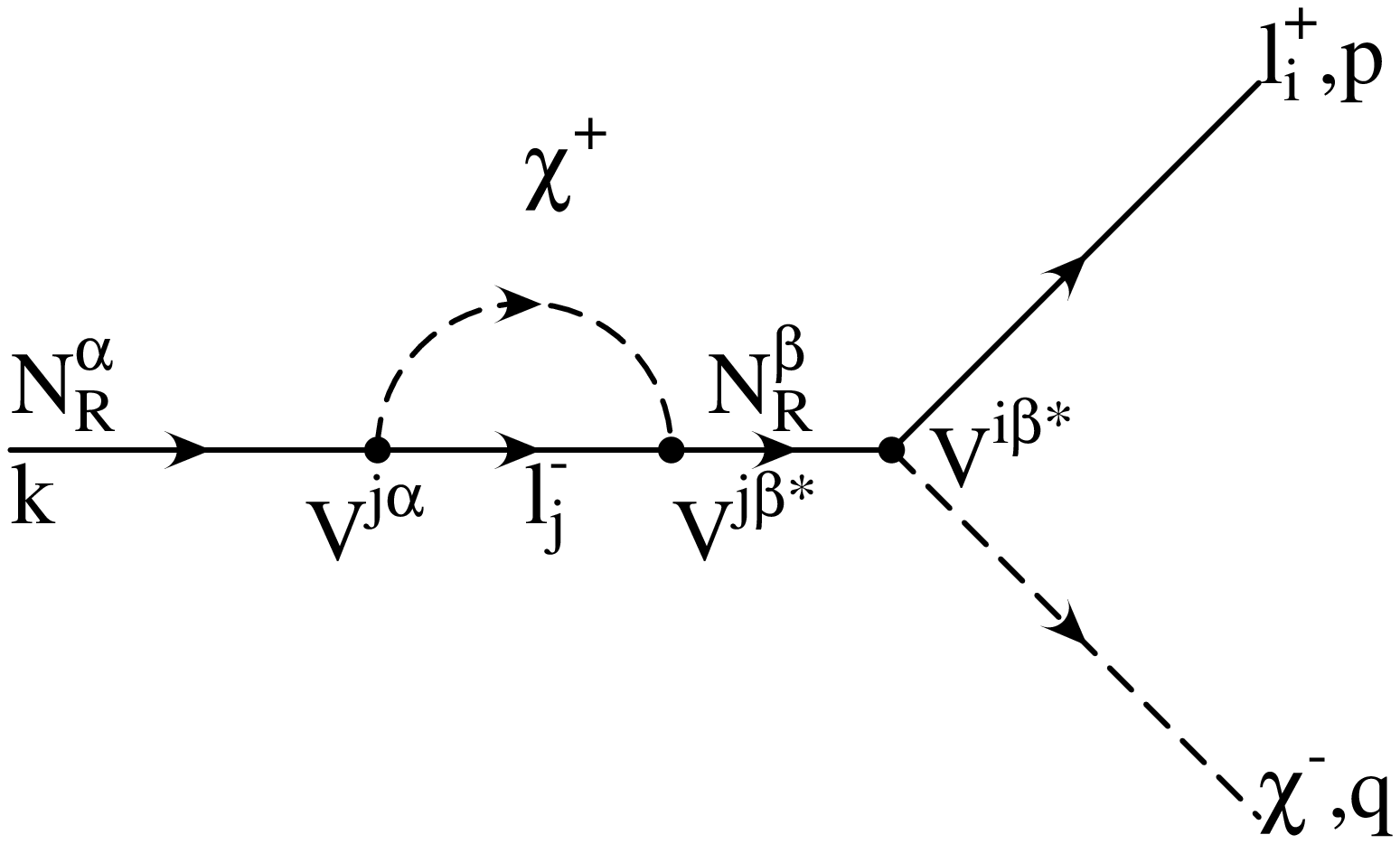,width=\hsize}
\label{fig:self-energy}
\end{center}
\end{minipage}
\end{tabular}
\caption{The diagrams of the right-handed neutrino Majorana
decay, i.e. tree, vertex and self energy diagram.}
\end{figure}
\begin{equation}
a_{{\alpha}i}^{v}=\sum_{j{\beta}}\frac{1}{8{\pi}}\Biggl(\frac{-\sqrt{2}}{v}m_{\beta}\Biggr)^{2}\mbox{Im}\Bigl[V_{MNS}^{i{\beta}}V_{MNS}^{j{\beta}}(V_{MNS}^{j{\alpha}})^{*}(V_{MNS}^{i{\alpha}})^{*}\Bigr]I\Biggl(\frac{m_{\beta}^{2}}{m_{\alpha}^{2}}\Biggr)\frac{1}{\sum_{i}|V_{MNS}^{i{\alpha}}|^{2}}. \label{19}
\end{equation}
Similarly, the interference between tree and self-energy diagram gives
the asymmetry 
$a_{\alpha i}^{s}$ as

\begin{equation}
a_{\alpha i}^{s} =\sum_{j{\beta}}\frac{1}{16{\pi}}\Biggl(\frac{-\sqrt{2}m_{\beta}}{v}\Biggr)^{2}\frac{m_{\alpha}^{2}}{m_{\alpha}^{2}-m_{\beta}^{2}}\mbox{Im}\Bigl\{V_{MNS}^{i{\beta}}V_{MNS}^{j{\beta}}(V_{MNS}^{j{\alpha}})^{*}(V_{MNS}^{i{\alpha}})^{*}\Bigr\}\frac{1}{\sum_{i}|V_{MNS}^{i{\alpha}}|^{2}}. 
\label{20}
\end{equation} 
The asymmetry in question is summation Eqs.(\ref{19}) and (\ref{20}),
\begin{eqnarray}
&\ &a(N_{\alpha}{\rightarrow} l_i^{\mp}{\chi}^{\pm}) \nonumber \\
&=& \sum_{j{\beta}}\frac{1}{8{\pi}}\Biggl(\frac{-\sqrt{2}}{v}m_{\beta}\Biggr)^{2}\mbox{Im}\Bigl[V_{MNS}^{i{\beta}}V_{MNS}^{j{\beta}}(V_{MNS}^{j{\alpha}})^{*}(V_{MNS}^{i{\alpha}})^{*}\Bigr]\frac{1}{\sum_{i}|V_{MNS}^{i{\alpha}}|^{2}} \nonumber \\
&~& {\times}\Biggl\{I\Biggl(\frac{m_{\beta}^{2}}{m_{\alpha}^{2}}\Biggr)+\frac{1}{2}\frac{m_{\alpha}^{2}}{m_{\alpha}^{2}-m_{\beta}^{2}}\Biggr\}. 
\label{21}
\end{eqnarray}

From FIG.1, ${\alpha}=4,5,6$ is index for decaying massive right-handed
neutrinos. If we write it with the new index ${l}{\equiv}\alpha-3$,
and taking into account the fact that the light neutrinos have mass
much smaller than heavy ones as well as the vacuum expectation value
of Higgs field $v$, then the contribution to the asymmetry effectively comes from the heavy part
only. Substituting the heavy part of $V_{MNS}$; of Eq.(\ref{8}) into (\ref{21}) we obtain 
\begin{equation}
a_{li}=\frac{1}{8{\pi}}\sum_{jk}\mbox{Im}\Bigl[Y^{ik}Y^{jk}(Y^{jl})^{*}(Y^{il})^{*}\Bigr]\frac{1}{\sum_{i}|Y^{il}|^{2}}\Biggl\{I\Biggl(\frac{m_{Nk}^{2}}{m_{Nl}^{2}}\Biggr)+\frac{1}{2}\frac{m_{Nl}^{2}}{m_{Nl}^{2}-m_{Nk}^{2}}\Biggr\}. \label{22}
\end{equation} 
It is clear that the heavy phase appear in lepton number asymmetry at very
early of the universe. Summing up to the final charged lepton and
expressed in $Y_{\triangle}$ yields 
\begin{equation}
a(N_l{\rightarrow}l^{\mp}{\chi}^{\pm})=\sum_{i}a_{li}=\frac{1}{8{\pi}(Y^{\dagger}_{\triangle}Y_{\triangle})^{kk}}\sum_{k}\mbox{Im}\Bigl[(Y^{\dagger}_{\triangle}Y_{\triangle})^{lk}\Bigr]^{2}\Biggl\{I\Biggl(\frac{m_{Nk}^{2}}{m_{Nl}^{2}}\Biggr)+\frac{1}{2}\frac{m_{Nl}^{2}}{m_{Nl}^{2}-m_{Nk}^{2}}\Biggr\}. 
\label{23}
\end{equation} 
Assuming the off-diagonal elements of $Y_{\triangle}$ are much smaller
than the diagonal ones then the explicit form of the asymmetry (\ref{23})
given as follows
\begin{eqnarray}
a(N_1\ {\rightarrow}\ l^{\mp}{\chi}^{\pm})&=& \frac{-1}{8{\pi}y^{2}_{1}}\Bigl[y^2_{2} \mbox{Im}(y^2_{21})\Biggl\{I\Biggl(\frac{m_{N2}^{2}}{m_{N1}^{2}}\Biggr)+
\frac{1}{2}\frac{m_{N1}^{2}}{m_{N1}^{2}-m_{N2}^{2}}\Biggr\} \nonumber \\
&~&~~~~~+ y^2_{3} \mbox{Im}(y^2_{31})\Biggl\{I\Biggl(\frac{m_{N3}^{2}}{m_{N1}^{2}}\Biggr)+\frac{1}{2}\frac{m_{N1}^{2}}{m_{N1}^{2}-m_{N3}^{3}}\Biggr\}\Bigr].
\label{24}
\end{eqnarray} 
\begin{eqnarray}
a(N_2\ {\rightarrow}\ l^{\mp}{\chi}^{\pm})&=& \frac{1}{8{\pi}y^{2}_{2}}\Bigl[y^2_{2} \mbox{Im}(y^2_{21})\Biggl\{I\Biggl(\frac{m_{N1}^{2}}{m_{N2}^{2}}\Biggr)+\frac{1}{2}\frac{m_{N_{2}}^{2}}{m_{N2}^{2}-m_{N_{1}}^{2}}\Biggr\}
\nonumber \\ 
&~&~~~~~- y^2_{3} \mbox{Im}(y^2_{32})\Biggl\{I\Biggl(\frac{m_{N3}^{2}}{m_{N2}^{2}}\Biggr)+\frac{1}{2}\frac{m_{N2}^{2}}{m_{N2}^{2}-m_{N3}^{3}}\Biggr\}\Bigr]. 
\label{25}
\end{eqnarray} 
\begin{eqnarray}
a(N_3\ {\rightarrow}\ l^{\mp}{\chi}^{\pm})&=& \frac{1}{8{\pi}}\Bigl[\mbox{Im}(y^2_{31})\Biggl\{I\Biggl(\frac{m_{N1}^{2}}{m_{N3}^{2}}\Biggr)+\frac{1}{2}\frac{m_{N_{3}}^{2}}{m_{N_{3}}^{2}-m_{N1}^{2}}\Biggr\}
\nonumber \\
&~&~~~~~+ \mbox{Im}(y^2_{32})\Biggl\{I\Biggl(\frac{m_{N2}^{2}}{m_{N3}^{2}}\Biggr)+\frac{1}{2}\frac{m_{N_{3}}^{2}}{m_{N3}^{2}-m_{N_{2}}^{2}}\Biggr\}\Bigr]. 
\label{26}
\end{eqnarray}
From the above results we can see that the leptogenesis does not depend
on the Dirac and Majorana phases, and depend only on the phase in the 
off-diagonal Yukawa coupling $y_{ij}$.

To see more explicitly the effects of Yukawa coupling we describe it
in geometrical form below. The effects can be seen from the neutral
current $j_{NC}^{\mu}=
\overline{\nu}_{L}^{\alpha}{\gamma}^{\mu}Z^{{\alpha}{\beta}}{\nu}_{L}^{\beta}$,
where the explicit form of $Z^{{\alpha}{\beta}}$ is given in
Appendix B. Tree level Z FCNC in the neutrino sector occurs, and the
FCNC comes out at $O({v^2}/{m^{2}_N})$. In general, in this geometrical
representation, the effect of $Z$ is to open the closed
polygon as given in decoupling case by unitarity mixing matrix. We consider two generation case, $Z_{12}$ is
given as follows
\begin{equation}
Z_{12}{\simeq}\frac{m_{{\nu}_{1}}m_{{\nu}_{2}}}{v^{2}}\frac{y_{21}}{y_{1}y_{2}^{2}}. 
\label{27}
\end{equation} 
The effect of $Z_{12}$ is to open two lines which are very close
together as consequence of unitarity of two generation case and shown in FIG.2.

\begin{figure}[t]
\begin{minipage}{0.7\hsize}
\begin{center}
\epsfig{file=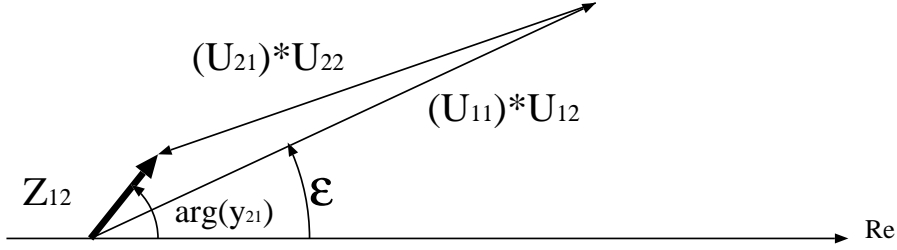,width=\hsize}
\caption{\small Effect of lepton number asymmetry (2 generation case)}
\label{fig:triangle}
\end{center}
\end{minipage}
\end{figure}

We consider now the three generation case. The $Z_{ij}$'s are given
\begin{equation}
\begin{array}{l}
{\displaystyle Z^{12}{\simeq} -\frac{m_{{\nu}_{1}}m_{{\nu}_{2}}}{v^{2}}\frac{y_{21}}{y_{1}y_{2}^{2}}}. \\
{\displaystyle Z^{13}{\simeq} -\frac{m_{{\nu}_{1}}m_{{\nu}_{3}}}{v^{2}}\frac{y_{31}}{y_{1}y_{3}^{2}}}. \\
{\displaystyle Z^{23}{\simeq} -\frac{m_{{\nu}_{2}}m_{{\nu}_{3}}}{v^{2}}\frac{y_{32}}{y_{2}y_{3}^{2}}}.
\end{array}
\label{28}  
\end{equation}
In both cases i.e. two and three generation case we have assumed that
the diagonal elements of Yukawa matrix are much bigger than the
off-diagonal ones. $Z_{ij}$'s depend on the off-diagonal elements of
$Y_{\triangle}$. $Z_{ij}$ has length and phase, and the effect of this
quantity, for example in three generation case, is to open the closed
unitarity triangle in decoupling case as shown in FIG.3.

\begin{figure}[t]
\begin{tabular}{cc}
\begin{minipage}{0.5\hsize}
\begin{center}
\epsfig{file=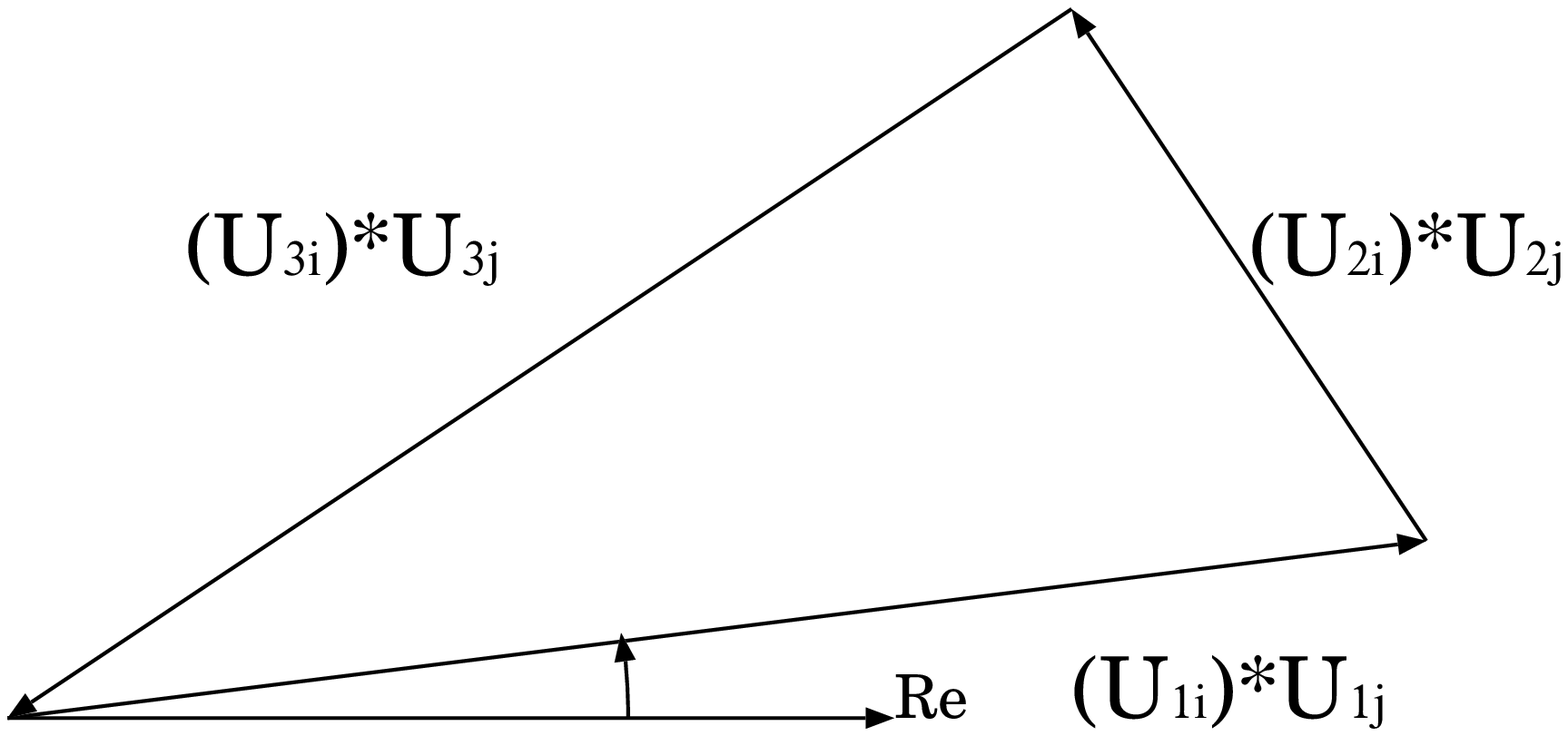,width=\hsize}
\label{fig:triangle}
\end{center}
\end{minipage}
\begin{minipage}{0.5\hsize}
\begin{center}
\epsfig{file=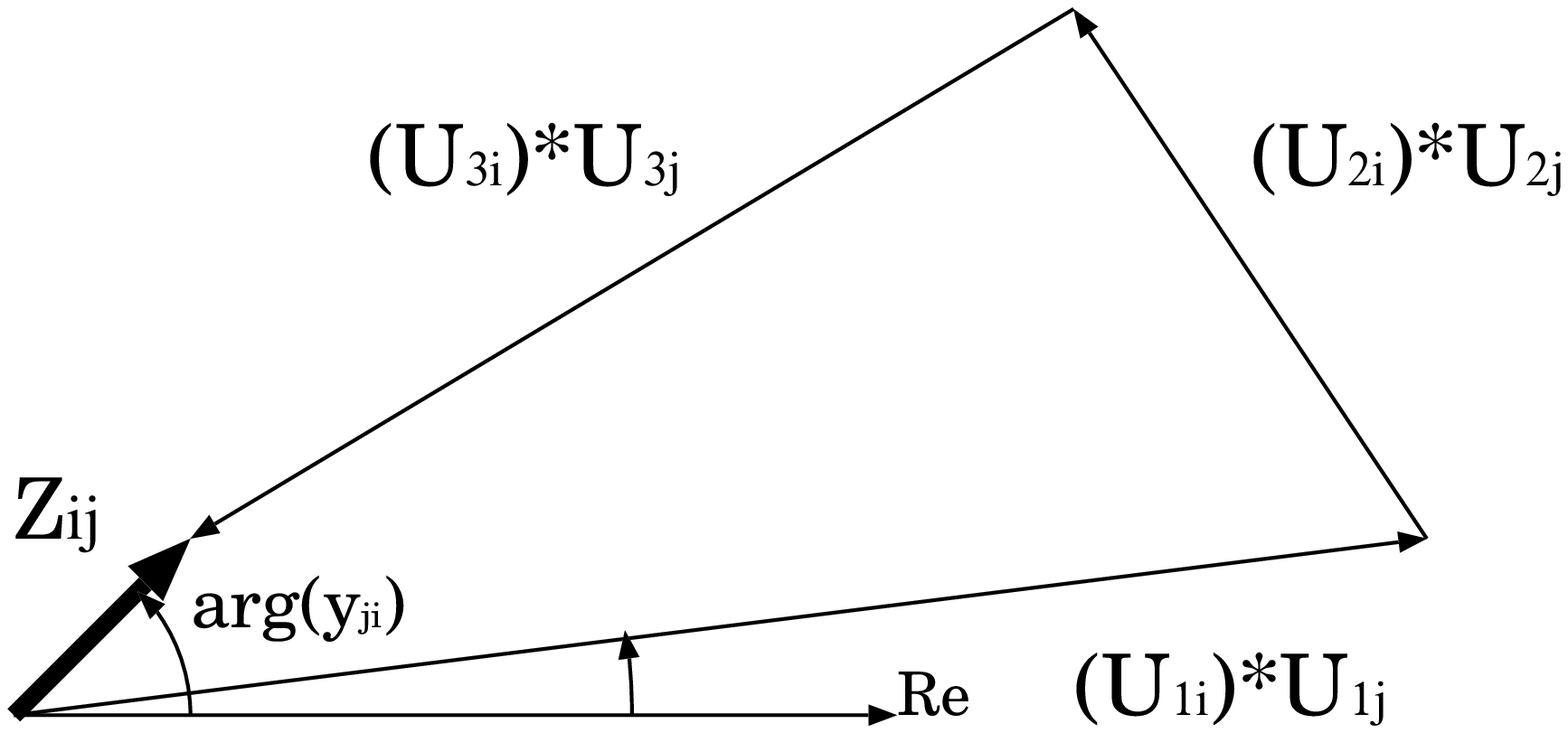,width=\hsize}
\label{fig:quatrangle}
\end{center}
\end{minipage}
\end{tabular}
\caption{\small Unitarity ``closed'' and ``open'' triangle.}
\end{figure}

\section{Discussion and Conclusion}
We studied the structure of CP phase of the seesaw model and found
that there are six independent phases rather than three phases as in Majorana
neutrino. The extra three phases come from Yukawa coupling between
light left-handed neutrino and heavy right-handed neutrino. From the
six phases, one, two and three phase are sensitive in neutrino
oscillation, neutrinoless double beta decay and lepton number asymmetry
respectively. The effect of heavy part hidden in Yukawa coupling -in
geometrical representation- open the unitary triangle of low energy sector. 
Our analysis on the experimental observation of the CP violation 
and its geometrical interpretation is based on a specific 
case with hierarchy($m_{N1} \gg m_{N2} \gg m_{N3} \gg v$) where
the explicit parameterization by the triangle method is possible.
In the present case, we found that the three experimental observations
give almost independent information on the CP violating phases.
It would be important to see whether this is also the case in 
more general cases. It would also be interesting to extend our 
geometrical interpretation to models with sterile fermions where 
FCNC give non-negligible effects.

\section {Acknowledgement}
The authors would like to thank K.~Funakubo for useful discussions. 
T.~M. is supported by the Grants-in-Aid for Scientific Research 
on Priority Areas ( Physics of CP violation, No. 12014211 ) and by
the Grand-in-Aid for JSPS Fellows ( No. 98362 ).
T.~O. is supported by the Grants-in-Aid of the Ministry of 
Education ( No. 12640279 ).
A.~P. would like to thank Ministry of
Education and Culture of Japan for the financial support under Monbusho
Fellowship Program. T. E. and A. P. thank Post Summer Institute 2000
(PSI2000) where a part of this work has been done.

\appendix
\section{Triangle method} 

Here we will show the relation of the phases between the unitary
matrix $U$, Yukawa coupling $Y_{\nu}$ and the triangle form of Yukawa
coupling $Y_{\triangle}$, where the theorem for the relation has been
given in~\cite{MoTa}. 
For general Yukawa coupling matrix $Y_{\nu}$ we can always find
unitary matrix $U$ which makes it becomes triangular matrix as in
Eq.(\ref{9}). To simplify we write $Y_{\nu}$ in the form
$$
Y_{\nu}=\left(\begin{array}{ccc}
f_{1} & g_{1} & z_{1}e^{i{\alpha}_{1}} \\
f_{2} & g_{2} & z_{2}e^{i{\alpha}_{2}} \\
f_{3} & g_{3} & z_{3}e^{i{\alpha}_{3}} 
\end{array}\right),\eqno(A.1) \label{A1}
$$
where $f_{i}$ and $g_{i}$ are complex, whereas $z_{i}$ is
real. To do it we have two steps, first introduce a unitary matrix T. Writing $Y_{\nu}$ in the form Eq.(A.1) lead to the form of unitary matrix T
$$
T=\left(\begin{array}{ccc}
u_{1} & v_{1} & \frac{z_{1}}{z}e^{i{\alpha}_{1}} \\
u_{2} & v_{2} & \frac{z_{2}}{z}e^{i{\alpha}_{2}} \\
u_{3} & v_{3} & \frac{z_{3}}{z}e^{i{\alpha}_{3}} 
\end{array}\right){\equiv}({\bf u}\ {\bf v}\ {\bf w}), \eqno(A.2) \label{A2}
$$
where $z=\sqrt{z_{1}^{2}+z_{2}^{2}+z_{3}^{2}}$, and the multiplication
$T^{\dagger}Y_{\nu}$ yields
$$
T^{\dagger}Y_{\nu}= \left(\begin{array}{ccc}
{\bf u}^{\dagger}{\bf f} & {\bf u}^{\dagger}{\bf g} & 0 \\
{\bf v}^{\dagger}{\bf f} & {\bf v}^{\dagger}{\bf g} & 0 \\
{\bf w}^{\dagger}{\bf f} & {\bf w}^{\dagger}{\bf g} & z
\end{array}\right). \eqno(A.3) \label{A3}
$$
Eq.(A.3) and unitarity condition of $T$ lead to the result for T as follows
$$
T=\left(\begin{array}{ccc}
e^{i{\alpha}_{1}} & \ & \ \\
\ & e^{i{\alpha}_{2}} & \ \\
\ & \ & e^{i{\alpha}_{3}} 
\end{array}\right)\left(\begin{array}{ccc}
 -\cos{\phi}_{1} & \ 0 & \sin{\phi}_{1} \\
\sin{\phi}_{1}\sin{\phi}_{2}&-\cos{\phi}_{2} & \cos{\phi}_{1}\sin{\phi}_{2} \\
\sin{\phi}_{1}\cos{\phi}_{2}& \sin{\phi}_{2} & \cos{\phi}_{1}\cos{\phi}_{2}
\end{array}\right)\left(\begin{array}{ccc}
e^{i{\alpha}_{4}} & \ & \ \\
\ & e^{i{\alpha}_{5}} & \ \\
\ & \ & 1 
\end{array}\right).\eqno(A.4)  \label{A4}
$$
having two angles, $({\phi}_{1},{\phi}_{2})$ and five phases, $({\alpha}_{1},{\cdots},{\alpha}_{5})$.

The second step is introducing another unitary matrix $S$. In this
step the present procedure is simpler than the same step
in~\cite{MoTa}. We do the similar procedure with one in the first step
i.e. to make zero the element $[S^{\dagger}T^{\dagger}Y_{\nu}]_{12}$,
whereas in the later using Gram-Schmidt diagonalization and then
become little complicated. For our requirement we rewrite
$T^{\dagger}Y$ into the form
$$
T^{\dagger}Y_{\nu}
{\equiv}\left(\begin{array}{ccc}
a_{1} & b_{1}e^{i{\sigma}_{1}} & 0 \\
a_{2} & b_{2}e^{i{\sigma}_{2}} & 0 \\
{\bf w}^{\dagger}{\bf f} & {\bf w}^{\dagger}{\bf g} & z
\end{array}\right)=\left(\begin{array}{ccc}
{\bf A} & {\bf B} & {\bf 0} \\ 
{\bf w^{\dagger}f} & {\bf w^{\dagger}g} & z
\end{array}\right). \eqno(A.5) \label{A5}
$$
where $a_{i}$ is complex, and $b_{i}$ is real. The above form of
$T^{\dagger}Y_{\nu}$ leads to the form of $S$ 
$$
S= \left(\begin{array}{ccc}
{\bf s} & \frac{{\bf B}}{|{\bf B}|} & {\bf 0} \\ 
0 & 0 & 0
\end{array}\right)= \left(\begin{array}{ccc}
s_{11} & \sin{\phi}_{3}e^{i{\sigma}_{1}} & 0 \\
s_{21} & \cos{\phi}_{3}e^{i{\sigma}_{2}} & 0 \\
0 & 0 & 0
\end{array}
\right), \eqno(A.6)\label{A6}
$$
where ${\phi}_{3}=\arctan (b_{1}/b_{2})$. The requirement of the form
$[S^{\dagger}T^{\dagger}Y]_{12}$ is zero and the unitarity of $S$ lead
to the form 

$$
S= \left(\begin{array}{ccc}
e^{i{\sigma}_{1}} & 0 & 0 \\
0 & e^{i{\sigma}_{2}} & 0 \\
0 & 0 & 1 
\end{array}\right)\left(\begin{array}{ccc}
\cos{\phi}_{3} & \sin{\phi}_{3} & 0 \\
-\sin{\phi}_{3} & \cos{\phi}_{3} & 0 \\
0 & 0 & 1
\end{array}\right)\left(\begin{array}{ccc}
e^{i{\epsilon}} & \ \\
\ & 1 \\
\ & \ & 1
\end{array}\right). \eqno(A.7) \label{A7}
$$
where
${\epsilon}$ is a free parameter and can be chosen such that
$[S^{\dagger}(T^{\dagger}y)]_{11}$ becomes real, and hence, all
diagonal elements of $[S^{\dagger}T^{\dagger}Y]$ are real .
From Eqs.(A.3), (A.4) and (A.7) we obtain $U$ matrix
$$
U = TS= \underbrace{\underbrace{\left(\begin{array}{ccc}
e^{i{\alpha}_{1}} & \ & \ \\
\ & e^{i{\alpha}_{2}} & \ \\
\ & \ & e^{i{\alpha}_{3}}
\end{array}\right)}_{\tiny \mbox{3 phases}}\underbrace{\left(\begin{array}{ccc}
-c_{1} & 0 & s_{1} \\
s_{1}s_{2} &-c_{2} & c_{1}s_{2} \\
s_{1}c_{2} &s_{2} & c_{1}c_{2}
\end{array}\right)}_{\tiny \mbox{2 angles}}\underbrace{\left(\begin{array}{ccc}
e^{i{\beta}} & \ & \ \\
\ & e^{i{\gamma}} & \ \\
\ & \ & 1
\end{array}\right)}_{\tiny \mbox{2 phases}}\underbrace{\left(\begin{array}{ccc}
c_{3} & s_{3} & 0 \\
-s_{3} & c_{3} & 0 \\
0 & 0 & 1 
\end{array}\right)}_{\tiny \mbox{1 angle}}\underbrace{\left(\begin{array}{ccc}
e^{i{\epsilon}} & \ & \ \\
\ & 1 & \ \\
\ & \ & 1 
\end{array}\right)}_{\tiny \mbox{1 phase}}}_{\tiny \mbox{3 angles + 6
phases}} \eqno(A.8) \label{A8}
$$
where $c_i$, $s_j$ are $\cos{\phi}_i$, $\sin{\phi}_j$ and we have reduced four parameters in $T$ and $U$ and expressed as
two independent parameters ${\alpha}_{4}+{\sigma}_{1}{\rightarrow}{\beta}$
and ${\alpha}_{5}+{\sigma}_{2}{\rightarrow}{\gamma}$.

To see the content of independent parameter in $Y_{\nu}$, $U$ and
$Y_{\triangle}$ we consider Eq.(\ref{9}) 
$$
Y_{\nu} = {\bf U}\left(\begin{array}{ccc}
y_{1} & 0 & 0 \\
y_{21} & y_{2} & 0 \\
y_{31} & y_{32} & y_{3} 
\end{array}\right), \eqno(A.9) \label{A9}
$$
From the above expression we can see that the number of independent
parameter of $Y$ is the same with $UY_{\triangle}$, i.e. 18 parameters.
For the phase of the right term, nine parameter is in $U$ and the other nine in $Y_{\triangle}$. We have
seen that the diagonal elements of $Y_{\triangle}$ are real, then this
$Y_{\triangle}$ has three independent phases live in off-diagonal
elements. 

For the physical consideration, the phase ${\alpha}_i$ of U in
Eq.(A.8) may be absorbed by physical fields and U becomes $U'$,
$$
U' = \left(\begin{array}{ccc}
-c_{1} & 0 & s_{1} \\
s_{1}s_{2} &-c_{2} & c_{1}s_{2} \\
s_{1}c_{2} &s_{2} & c_{1}c_{2}
\end{array}\right)\left(\begin{array}{ccc}
e^{i{\beta}} & \ & \ \\
\ & e^{i{\gamma}} & \ \\
\ & \ & 1
\end{array}\right)\left(\begin{array}{ccc}
c_{3} & s_{3} & 0 \\
-s_{3} & c_{3} & 0 \\
0 & 0 & 1 
\end{array}\right)\left(\begin{array}{ccc}
e^{i{\epsilon}} & \ & \ \\
\ & 1 & \ \\
\ & \ & 1 
\end{array}\right) \eqno(A.10)\label{A10}
$$
After we rearrange the phases and omit irrelevant part we obtain 
$$
U'=\left(\begin{array}{ccc}
-c_{1} & 0 & s_{1} \\
s_{1}s_{2} &-c_{2} & c_{1}s_{2} \\
s_{1}c_{2} &s_{2} & c_{1}c_{2}
\end{array}\right)
\left(\begin{array}{ccc}
e^{i{\theta}} & \ & \ \\
\ & e^{-i{\theta}} & \ \\
\ & \ & 1 
\end{array} \right)\left(\begin{array}{ccc}
c_{3} & s_{3} & 0 \\
-s_{3} &c_{3} & 0 \\
0 & 0 & 1
\end{array}\right) \nonumber \\
\left(\begin{array}{ccc}
1 & \ & \ \\
\ & e^{-i{\zeta}_1} & \ \\
\ & \ & e^{-i{\zeta}_2} 
\end{array}\right), \eqno(A.11)\label{A11}
$$
where ${\theta}{\rightarrow} \frac{{\beta}-{\gamma}}{2}$ we identify
as Dirac phase, whereas
${\zeta}_{1}\ {\rightarrow}{\epsilon}$ and
$ {\zeta}_{2} {\rightarrow}\frac{{\beta}+{\gamma}}{2}+{\epsilon}$ as
Majorana phases.

\section{FCNC}
Consider the neutral current
$$
j_{NC}^{\mu}= \overline{\nu}_{L}^{0i}{\gamma}^{\mu}{\nu}_{L}^{0i} = \overline{\nu}_{L}^{\alpha}{\gamma}^{\mu}Z^{{\alpha}{\beta}}{\nu}_{L}^{\beta} \eqno(B.1)\label{B1}
$$
where $Z^{{\alpha}{\beta}}$ as in ~\cite{KiTa} is given
$$
Z^{{\alpha}{\beta}}= (V^{\dagger})^{{\alpha}i}V^{i{\beta}} = {\delta}^{{\alpha}{\beta}}-(V^{I{\alpha}})^{*}V^{I{\beta}}, \eqno(B.2)\label{B2}
$$
the explicit form of $V$ is given as follows~\cite{HaMo}
$$
V=\left(\begin{array}{cc}
U & U{Y_{\triangle}}^{*}\frac{v}{\sqrt{2}m_{N}} \\
-\frac{v}{\sqrt{2}m_{N}}{Y_{\triangle}}^{T} & 1 
\end{array}\right). \eqno(B.3)\label{B3}
$$
For $1{\leq}{\alpha}{\neq}{\beta}{\leq}3$, we obtain
$$
 Z^{ij}= 
 -(V^{4i})^{*}V^{4j}-(V^{5i})^{*}V^{5j}-(V^{6i})^{*}V^{6j}.  \eqno(B.4)\label{B4}
$$
In non-decoupling
case $Z^{ij}$ is not zero then it give effect -for three generation
case- on the triangle and
becomes open triangle.

Substitute the explicit form Eq.(B.3) into Eq.(B.4) yields
$$
-Z_{ij}
= \sum_{k=1}^{3}(Y_{\triangle})^{ik}\frac{v^{2}}{2m_{Nk}}(Y_{\triangle}^{\dagger})^{kj}, \eqno(B.5)\label{B5}
$$
and then using relation of left-handed Majorana neutrino mass~\cite{MoTa}
$$
m_{\nu}=Y_{\triangle}\frac{v^{2}}{{\sqrt{2}}m_{N}}Y_{\triangle}^{T}, \eqno(B.6)\label{B6}
$$
we obtain
$$
-Z_{ij}= \frac{m_{{\nu}i}m_{{\nu}j}}{v^{2}}\Bigl[(Y_{\triangle}^{*}Y_{\triangle}^{T})^{-1}\Bigr]_{ij}. \eqno(B.7)\label{B7} 
$$
Geometrical representation of this formulation for
two generation case is given by the mixing matrix $U$
$$
U= \left(\begin{array}{cc}
\cos{\theta} & -\sin{\theta}e^{i{\epsilon}} \\
\sin{\theta}e^{-i{\epsilon}} & \cos{\theta}
\end{array}\right). \eqno(B.8)\label{B8}
$$
and the triangular Yukawa coupling matrix
$$
Y_{\triangle}=\left(\begin{array}{cc}
y_{1} & 0 \\
y_{21} & y_{2}
\end{array}
\right). \eqno(B.9)\label{B9}
$$
the result is shown by FIG.2. and Eq.(\ref{27}). For three generation
case, the unitarity triangle of $U$ in decoupling case are open by
$Z_{ij}'s$ given in Eq.(\ref{28}) as shown by FIG.3., where $Y_{\triangle}$ is given in Eq.(\ref{9}).


\end{document}